\documentclass[twocolumn,pra,english,aps,floatfix,superscriptaddress,showpacs,amsfonts,amssymb,superscriptaddress,nofootinbib]{revtex4}

\usepackage{color} 
\usepackage{hyperref}
\usepackage{slashed}
\usepackage{graphicx}
\usepackage{amsmath}
\usepackage{latexsym}
\usepackage{epstopdf}
\usepackage{amsmath}
\usepackage{enumitem}
\usepackage{amssymb,amsmath}
\usepackage{multirow}
\usepackage{mathtools}
\usepackage{bbm}
\usepackage{bm}
\usepackage{amsfonts}
\usepackage{amssymb}
\usepackage{calligra}
\usepackage{calrsfs}
\usepackage{makecell}

\newcommand{\beq}{\begin{equation}}
\newcommand{\eeq}{\end{equation}}

\def\ket#1{|#1\rangle}
\newcommand{\mean}[1]{\ensuremath{\langle #1 \rangle}}
\newcommand{\Eins}{\ensuremath{\mathbbm 1}}

\newcommand{\be}{\begin{equation}}
\newcommand{\ee}{\end{equation}}
\newcommand{\xiR}{\xi_{\rm R}}

\begin{document}

%\title{Loschmidt Echo Quantum Metrology with Rydberg Lattices}
\title{Loschmidt echo for quantum metrology}

\author{Tommaso Macr\`i}
\affiliation{Departamento de F\'isica Te\'orica e Experimental, Universidade Federal do Rio Grande do Norte, 
and International Institute of Physics, Natal-RN, Brazil}
\author{Augusto Smerzi}
\author{Luca Pezz\`e}
\affiliation{QSTAR, INO-CNR and LENS, Largo Enrico Fermi 6, I-50125, Firenze, Italy}
\date{\today{}}

\begin{abstract}
%Entanglement is a key resource for applications to fundamental physics as well as to quantum technologies.
%The characterization of multiparticle entangled states is usually a difficult task both on theoretical and experimental side. 
%Multiparticle entanglement is related, via the quantum Fisher information, to the phase sensitivity in quantum metrology. 
We propose a versatile Loschmidt echo protocol to detect and quantify multiparticle entanglement. %, with direct application to quantum metrology. 
It allows us to extract the quantum Fisher information for arbitrary pure states, and finds direct application in quantum metrology. 
In particular, the protocol applies to states that are generally difficult to characterize, as 
non-Gaussian states, and states that are not symmetric under particle exchange.
We focus on atomic systems, including trapped ions, polar molecules, and Rydberg atoms, where entanglement is generated dynamically via long range interaction,
%ultracold atoms in deep optical lattices 
%excited to two different Rydberg levels, 
%We finally discuss a possible implementation with ultracold atoms in deep optical lattices 
%excited to two different Rydberg levels, %that display soft-core interactions with opposite sign, 
%Finally we characterize the quantum Fisher Information via a mapping to a spin-$1/2$ long-range lattice model that is amenable 
%to an exact analytic evaluation 
and show that the protocol is stable against experimental detection errors.
\end{abstract}
\maketitle

Engineering and detecting entangled states of many atoms is a vivid area of research~\cite{GuhnePHYSREP2009}.
Besides the intrinsic foundational interest, entangled states can find important technological 
applications in quantum metrology~\cite{GiovannettiNATPHOT2011, TothJPA2014, PezzeVARENNA2014}.
Most of the investigations and experimental protocols focus on Gaussian spin-squeezed states~\cite{MaPHYSREP2011}.
The generation of entangled non-Gaussian ({\it i.e.}, not spin-squeezed) states (ENGSs) of many atoms 
has been tackled only recently~\cite{StrobelSCIENCE2014, LuckeSCIENCE2011, Bohnet_2015, BarontiniSCIENCE2015, McConnellNATURE2015, HaasSCIENCE2011}.
Interestingly, in several cases, ENGSs outperform the metrological sensitivity achievable using spin-squeezed states created with the same entanglement-generation protocol. 
A prominent example is the dynamical evolution of a separable state of many qubits via long-range interaction in an Ising models, as described below.
How to detect and use those states?

Spin-squeezed states are fully characterized by mean values and variances of collective spin operators, and 
there are well known relations that link these quantities to entanglement~\cite{SorensenNATURE2001, TothPRL2007, MaPHYSREP2011, LuckePRL2014}.
For instance, metrological spin squeezing  $\xiR^2  = N(\Delta \hat{J}_{n_3})^2 / \mean{\hat{J}_{n_1}}^2 <1$~\cite{WinelandPRA1994}, 
where $\hat{J}_{n_i}$ is a collective spin operator, $N$ is the number of qubits, and $n_1$, $n_2$ and $n_3$ are three orthogonal directions, 
is a sufficient condition for particle entanglement~\cite{SorensenNATURE2001}.
By applying the transformation $e^{-i \theta \hat{J}_{n_2}}$, spin-squeezed states can be used for the estimation of the rotation angle $\theta$.
Looking at the mean spin as a function of $\theta$, 
it is possible to achieve a phase sensitivity $\Delta \theta = \xiR/\sqrt{N}$~\cite{WinelandPRA1994} that, 
when $\xiR<1$, is below the standard quantum limit $\Delta \theta_{\rm SQL} = 1/\sqrt{N}$ which gives the maximum sensitivity attainable with separable states~\cite{PezzePRL2009, GiovannettiPRL2006}. 

ENGSs are more difficult to detect. 
A useful condition is the entanglement criterion $F_Q[\hat{\rho}, \hat{J}_{n_2}] > N$~\cite{PezzePRL2009}, where $\hat{\rho}$ is a general state, 
and $F_Q$ is the quantum Fisher information (QFI)~\cite{HelstromBOOK, BraunsteinPRL1994, PezzeVARENNA2014}. 
In general $F_Q \geq N/\xiR^2$~\cite{PezzePRL2009}: the inequality $F_Q > N$ may thus detect entangled states that are not spin-squeezed ({\it i.e.} $\xiR \geq 1$).
This criterion has been further extended for the detection of multiparticle entanglement \cite{HyllusPRA2012}.
The QFI is directly related to metrological sensitivity by the quantum Cramer-Rao bound (QCRB) $\Delta \theta_{\rm QCR}  = 1/\sqrt{F_Q}$, giving 
the maximum phase sensitivity, optimized over all possible estimators and measurement strategies~\cite{HelstromBOOK, BraunsteinPRL1994}. 
Yet, the characterization and use of ENGSs for metrological sensing is generally hindered by substructures or tails of the phase-dependent probability distribution.
Furthermore, ENGSs that are non-symmetric under particle-exchange are challenging to study even theoretically, due to the Hilbert space dimension, 
exponentially increasing with the number of particles.   

In this Rapid Communication we propose an experimentally feasible Loschmidt echo~\cite{nota}
protocol to characterize and exploit general quantum states (including non-symmetric and non-Gaussian, in particular) for metrological applications,
see also~\cite{DavisPRL2016, FrowisPRL2015, GabbrielliPRL2015, LinemannArXiv2016}.
The protocol starts with a state $\ket{\psi_{\rm inp}}$ of $N$ qubits. 
We take, for instance, the product of $N$ spin-up particles, $\ket{\psi_{\rm inp}}=\ket{\uparrow}^{\otimes N}$. 
Particle entanglement is created dynamically by applying a nonlocal unitary evolution $\hat{U}_1$. 
This is followed by a rotation $e^{-i \theta \hat{J}_n}$, where $n$ is an arbitrary spin direction, and 
a second nonlocal transformation $\hat{U}_2$, which provides the echo operation.
The probability that the output state after the full protocol, $\ket{\psi_{\rm out}} = \hat{U}_2 e^{-i \theta \hat{J}_n} \hat{U}_1 \ket{\psi_{\rm inp}}$, 
coincides (up to a global phase factor) with the initial one is  
$P_0(\theta) = \vert \langle \psi_{\rm out} \ket{\psi_{\rm inp}} \vert^2$ (this quantity is also indicated as ``fidelity'' in the literature on Loschmidt echo problems \cite{nota}). 
Under the time reversal condition $\hat{U}_2 \hat{U}_1 = \Eins$, a Taylor series expansion around $\theta= 0$ gives
\be \label{Eq.P0}
P_0(\theta) = 1 - \frac{\theta^2}{4} F_Q\big[\ket{\psi_1}, \hat{J}_n \big] + \mathcal{O}(\theta^4),
\ee
where $F_Q[\ket{\psi_1}, \hat{J}_n] = 4 (\Delta \hat{J}_n)^2 = 4(\mean{\hat{J}_n^2} - \mean{\hat{J}_n}^2)$ is the QFI of the state $\ket{\psi_1} = \hat{U}_1 \ket{\psi_{\rm inp}}$.
We argue that the projection over the state $\ket{\uparrow}^{\otimes N}$ can be realized experimentally with very high efficiency (we comment on this later). 
Equation~(\ref{Eq.P0}) reveals that the decrease of $P_0(\theta)$ for $\theta  \approx 0$ is directly related to the QFI, 
which in turns depends on multiparticle entanglement in the quantum state $\ket{\psi_1}$~\cite{PezzePRL2009, HyllusPRA2012, Pezze_2015}.
Furthermore, we can use the probability $P_0(\theta)$ as phase-sensing signal. Standard error propagation predicts 
\be \label{Eq.sensitivity}
(\Delta \theta)^2 = \frac{ (\Delta P_0)^2}{ (d P_0 /d \theta)^2} \bigg \vert_{\theta=0}= \frac{1}{F_Q[\ket{\psi_1}, \hat{J}_n ]},
\ee 
where $(\Delta P_0)^2 = P_0 (1 - P_0) = F_Q \theta^2/4 + \mathcal{O}(\theta^4)$ and $(d P_0 /d \theta)^2 = F_Q^2 \theta^2/4 + \mathcal{O}(\theta^4)$. 
{In the ideal case,}
the Loschmidt echo, followed by the projection over the probe state, 
realizes a protocol to saturate the QCRB.
This holds under general conditions:
the unitary operators $\hat{U}_{1,2}$ can be generated by an arbitrary nonlocal Hamiltonian $\hat{H}$ and the 
scheme does not require any knowledge or assumption on the quantum state.
In particular, as we will illustrate in the following, the Loschmidt echo protocol applies to
ENGSs created with Ising-type long-range interaction. 
Furthermore, Eqs.~(\ref{Eq.P0}) and (\ref{Eq.sensitivity}) can be straightforwardly extended to generic qudit system. 
A protocol analogous to the one discussed in this paper has been recently analyzed in~\cite{DavisPRL2016, FrowisPRL2015}, 
where it was shown that the Loschmidt echo makes phase estimation robust against detection noise, see also~\cite{LinemannArXiv2016,nota1}. 
However, the possibility to saturate the QCRB with arbitrary states was not discussed in these works.  
Moreover, Refs.~\cite{DavisPRL2016, FrowisPRL2015} have focused on spin-squeezed states while, as shown here, 
the protocol applies to arbitrary ENGSs as well.

Noise, for instance detection noise or phase noise during the rotation, or an imperfect implementation of the echo ($\hat{U}_1 \hat{U}_2 \neq \Eins$), 
prevents the perfect compensation between numerator and denominator in Eq.~(\ref{Eq.sensitivity}) that lead 
to the result $(\Delta \theta)^2 = 1/F_Q$ at $\theta=0$.
In presence of noise, $(\Delta \theta)^2=(\Delta P_0)^2 / (d P_0 /d \theta)^2$ reaches its minimum at a finite value of $\theta$, and 
saturates $1/\sqrt{F_Q}$ in the limit of vanishing noise.
If the transformations $\hat{U}_{1,2}$ are not unitary, and in particular the state before the rotation $e^{-i \theta \hat{J}_n}$ is not pure, then 
Eq.~(\ref{Eq.sensitivity}) still gives an upper bound to the best achievable sensitivity, and $(d P_0 /d \theta)^2/(\Delta P_0)^2$ gives a lower bound to the QFI. 
Therefore, $(d P_0 /d \theta)^2/(\Delta P_0)^2>N$ implies $F_Q>N$, and it is thus a condition for entanglement. Conditions for multiparticle entanglement can be found following~\cite{HyllusPRA2012}.

\begin{figure}[t!]
%\begin{center}
\centering
\includegraphics[width=\columnwidth]{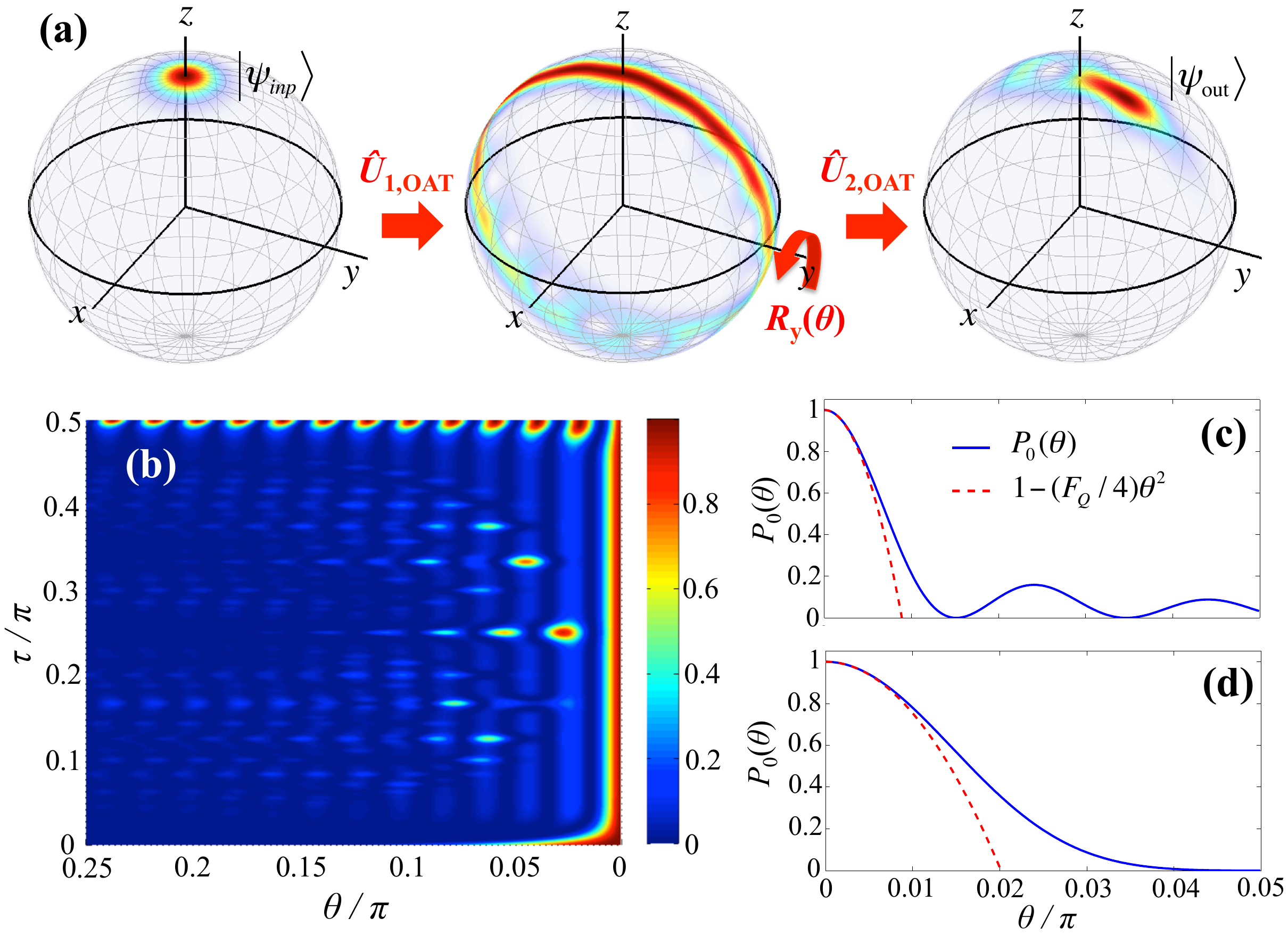}
\caption{(color online) 
Loschmidt echo protocol applied to the OAT model.
{(a)} Snapshot of the Husimi distribution.
(Left panel) A spin-polarized state is prepared at north pole of the Bloch sphere. 
(Central panel) Interaction is switched on for a time $t_1$ [transformation $\hat{U}_{1,{\rm OAT}}$].  
The state is then rotated around the $y$ axis of an angle $\theta$ [$\hat{R}_y(\theta)$]. 
(Right panel) Interaction is switched on again for a time $t_2$ [transformation $\hat{U}_{2,{\rm OAT}}$] such that $\hat{U}_{1,{\rm OAT}} \hat{U}_{2,{\rm OAT}} = \Eins$.  
In these plots $\theta/\pi=0.01$ and $\tau/\pi=0.05$.
{(b)} Probability $P_0(\theta)$ (color scale) as a function of time and phase shift. 
Panels {(c) and (d)} are cuts of panel (b) showing $P_0(\theta)$ (solid line) as a function of $\theta$ for $\tau/\pi=0.05$ (c) and $\tau/\pi=0.01$ (d).
The dashed line is the Taylor expansion as in Eq.~(\ref{Eq.P0}).
Here $N=100$.
}
\label{fig1}
%\end{center}
\end{figure}

As an example, we consider the Ising Hamiltonian
\be \label{LRI}
\hat{H} = \sum_{i,j=1}^N \frac{V_{ij}}{4} \hat \sigma_x^{(i)} \hat \sigma_x^{(j)}
\ee
where $\sigma_x^{(j)}$ is the Pauli matrix for the $j$th particle and 
$V_{ij}$ models the interaction strength between particles $i$ and $j$.
Our results are valid for arbitrary $V_{ij}$.
Hamiltonians of the type~(\ref{LRI}) have been experimentally implemented with power-law couplings $V_{ij} \approx V_0/(r_{ij}/a)^{\alpha}$, 
where $V_0$ is the on-site interaction strength (which can be tuned positive, or negative), $r_{ij}/a$ is the distance between particle $i$ 
and $j$ normalized to a characteristic distance, and $\alpha$ is a characteristic exponent: 
$0 \leq \alpha \leq 3$ for ions in a Penning trap~\cite{BrittonNATURE2012, Bohnet_2015},
$\alpha=3$ for polar molecules~\cite{GorshkovPRL2008,YanNATURE2013}, and $\alpha=6$ for Rydberg atoms trapped in an optical lattice 
\cite{Schauss2012,Zeiher2015,Labuhn2015,Zeiher2016,Schauss2015}.
When $V_{ij}=V_0$ we recover the one-axis twisting (OAT) model \cite{KitagawaPRA1993}, $\hat{H}_{\rm OAT} = V_0 \hat{J}_x^2$ 
with $\hat{J}_x = \tfrac{1}{2}\sum_{i=1}^N \sigma_x^{(i)}$.
OAT has been experimentally realized with Bose-Einstein condensates via atom-atom elastic collisions \cite{gross_2010, RiedelNATURE2010}, 
trapped ions~\cite{Bohnet_2015, MonzPRL2011} and, to a very good approximation, via off-resonance atom-light interaction in a optical cavity \cite{LerouxPRL2010}. 
The Loschmidt Echo protocol within the OAT model can be visualized in the Bloch sphere, as illustrated in Fig.~\ref{fig1}(a).
Starting with $\ket{\psi_{\rm inp}} = \ket{\uparrow}^{\otimes N}$,
particle entanglement is generated dynamically by applying $\hat{U}_{1,{\rm OAT}} = e^{-i (t V)_1 \hat{H}_{\rm OAT}}$ ($\hbar=1$ in the following), 
where $(V t)_1$ refers to the evolution for a time $t_1$ and an interaction strength $V_1$.
The state is then rotated around the $y$ axis of an angle $\theta$, $\hat{R}_y(\theta) = e^{-i \theta \hat{J}_y}$. 
The dynamics is finally inverted by applying $\hat{U}_{2,{\rm OAT}} = e^{+i (V t)_2 H_{\rm OAT}}$. 
For $(V t)_1=(V t)_2 = \tau$, the overlap $P_0(\theta)$ between the initial and the final state shows irregular oscillations as a function of $\theta$, see Fig.~\ref{fig1}(b)-(d), and 
Eq.~(\ref{Eq.P0}) holds for $\theta\approx 0$.
In particular, for $\tau=\pi/2$ we have $P_0(\theta) = \cos^2(N\theta/2)$.

%%%%%%%%%%%%%%%%%%%%%%%%%%%%%%%%%%%%%%%%%%%%%%%%%%%%%%%%
% figure 1
%%%%%%%%%%%%%%%%%%%%%%%%%%%%%%%%%%%%%%%%%%%%%%%%%%%%%%%%
%%%%%%%%%%%%%%%%%%%%%%%%%%%%%%%%%%%%%%%%%%%%%%%%%%%%%%%%%
%%%%%%%%%%%%%%%%%%%%%%%%%%%%%%%%%%%%%%%%%%%%%%%%%%%%%%%%%
%%%%%%%%%%%%%%%%%%%%%%%%%%%%%%%%%%%%%%%%%%%%%%%%%%%%%%%%%

\begin{figure}[t!] 
\includegraphics[width=0.48\textwidth]{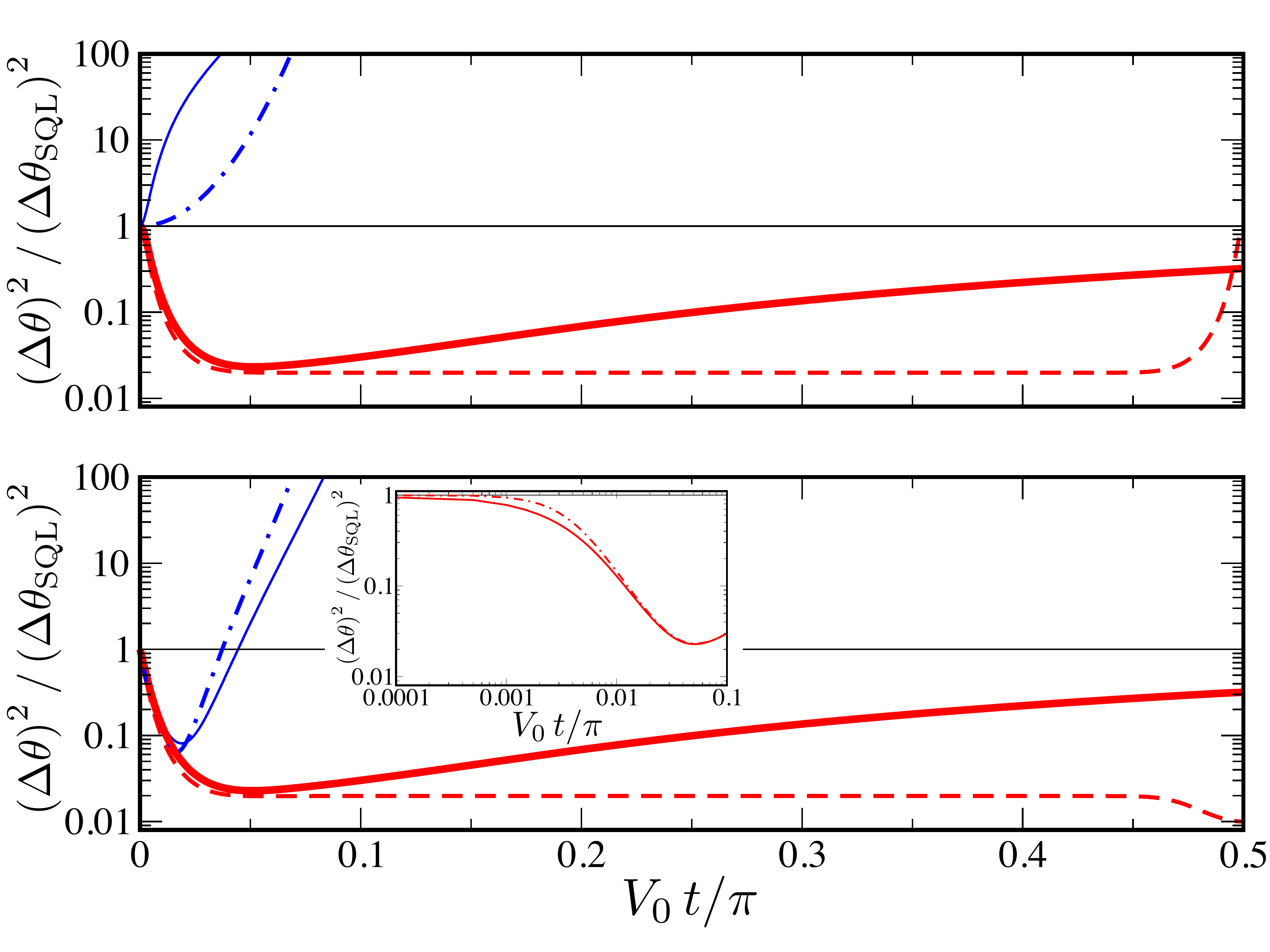}
\caption{(color online)
Phase sensitivity (normalized to the standard quantum limit) calculated as spin squeezing (thin solid and dashed-dotted blue lines) and inverse QFI (thick solid and dashed red lines).
Solid lines refer to a square lattice with $N=100$ particles with $L=10\, \text{a}_\text{latt}$ and $R_c=8 \, \text{a}_\text{latt}$, 
red dashed and blue dashed-dotted lines to the OAT model with $N=100$ particles.  
(a) Non-optimized case.
(b) Optimized case, where the state is rotated by a suitable angle before applying $R_y(\theta)$.
The inset shows the comparison of Eq.~(\ref{Eq.sensitivity}) for the optimized (solid) and non-optimized (dot-dashed) dynamics for the QFI at short times.}
\label{fig:dynamics}
\end{figure}

We can calculate average spin moments and variances of the state $\ket{\psi_1} = e^{-i t \hat{H}} \ket{\uparrow}^{\otimes N}$,
for arbitrary $V_{ij}$. 
These expectation values are used to compute the spin squeezing \cite{GilPRL2014}, and QFI.
We have $\mean{\hat{J}_x}= \mean{\hat{J}_y} = 0$ and 
\begin{equation}
\mean{\hat J_z}= \frac{1}{2}\sum_{i=1}^{N} \prod_{k\neq i}^N \cos(V_{ik} t), \nonumber
\end{equation}
as first moments, 
\begin{eqnarray} \label{collective_spins} \nonumber
\mean{ \hat J_x^2}  &=& \frac{N}{4} + \frac{1}{4}\sum_{i<j}^N
\bigg[ 
\prod_{k\neq i,j}^N \cos\left(\phi_{ijk}^{-}t \right) -
\prod_{k\neq i,j}^N \cos\left(\phi_{ijk}^{+}t \right)  
\bigg],\\ \nonumber
\mean{ \hat J_y^2 } &=& \frac{N}{4}, \\ \nonumber
\mean{ \hat J_z^2 } &=& \frac{N}{4} + \frac{1}{4}\sum_{i<j}^N
\bigg[ 
\prod_{k\neq i,j}^N \cos\left(\phi_{ijk}^{-}t \right) +
\prod_{k\neq i,j}^N \cos\left(\phi_{ijk}^{+}t \right)  \nonumber
\bigg],
\end{eqnarray}
where $\phi_{ijk}^{\pm} = V_{ik} \pm V_{jk}$, as second moments, and
\begin{equation}
\begin{array}{ccl}
\mean{ \hat J_x \hat J_z +\hat J_z \hat J_x } &=& \mean{ \hat J_y \hat J_z +\hat J_z \hat J_y } = 0, \\ \\
\mean{ \hat J_x \hat J_y +\hat J_y \hat J_x } &=&  \sum_{i<j}^N  \sin(V_{ij} t) \prod_{k\neq i,j} \cos(V_{ik}t). \nonumber
\end{array}
\end{equation}
As an example, we take a soft-core potential $V_{ij} = V_0 /[1+(r_{ij}/R_c)^6]$, 
where $R_c$  is the interaction range.
This potential is relevant for Rydberg dressed atoms~\cite{Henkel2010,macri_2014}, as we discuss below.
Due to non-uniform interactions, the state $\ket{\psi_1}$ is not restricted to the subspace of states symmetric under particle exchange.
In Fig.~\ref{fig:dynamics} we plot the phase sensitivity $(\Delta \theta)^2$ normalized to the standard quantum limit, 
where  $(\Delta \theta)^2 = (\Delta \theta)^2_{\rm CRB} = 1/F_Q[\ket{\psi_1}, \hat{J}_y]$ (solid red line), and 
$(\Delta \theta)^2 = \xiR^2/N$ (solid blue lines), where $\xiR^2 = N(\Delta \hat{J}_x)^2/\mean{\hat{J}_z}^2$, as a function of the evolution time.
The calculation is done assuming a uniform unit-filling two-dimensional lattice with $100$ atoms 
and $R_c=8\times \text{a}_{\text{latt}}$, where $a_\text{latt}$ is the lattice spacing.
In Fig.~\ref{fig:dynamics}(b) the spin squeezing and the QFI are optimized over all possible spin directions.
Notably, such optimization is crucial to achieve spin squeezing~\cite{GilPRL2014}. 
On the contrary, the QFI in the optimized and non-optimized cases differ only slightly and for relatively short time [see inset of Fig.~\ref{fig:dynamics}(b)]. 
After a transient time, spin squeezing is lost ($\xiR^2 \geq 1$) and ENGSs are produced. 
Minima of the inverse QFI and squeezing are both obtained at times much shorter
than typical interaction times ($V_0\, t \sim 1$).

%%%%%%%%%%%%%%%%%%%%%%%%%%%%%%%%%%%%%%%%%%%%%%%%%%%%%%%%
% figure 2
%%%%%%%%%%%%%%%%%%%%%%%%%%%%%%%%%%%%%%%%%%%%%%%%%%%%%%%%
%%%%%%%%%%%%%%%%%%%%%%%%%%%%%%%%%%%%%%%%%%%%%%%%%%%%%%%%%

The entanglement created dynamically strongly depends on the blockade radius compared to the typical system size.
For systems smaller than the characteristic interaction range the dynamics can be mapped into the OAT-type Hamiltonian.
spin squeezing~\cite{KitagawaPRA1993, SorensenNATURE2001} and QFI~\cite{PezzePRL2009} can be calculated analytically. 
We have
\be \label{FQ_OAT_1}
F_Q\big[\ket{\psi_1}, \hat{J}_y \big]  = \frac{N(N+1)}{2} - \frac{N(N-1)}{2} (\cos 2 V_0 t)^{N-2},
\ee
that predicts a phase sensitivity overcoming the standard quantum limit at any time for which $(\cos 2 V_0 t)^{N-2} \neq 1$. 
Since $(\cos 2 V_0 t)^{N-2} \approx e^{-2(N-2) V_0^2 t^2}$ for $N \gg 1$, the second term in Eq.~(\ref{FQ_OAT_1}) vanishes for $1/\sqrt{N} \lesssim V_0 t \lesssim \pi/2 - 1/\sqrt{N}$.
In this regime Eq. (\ref{Eq.sensitivity}) reaches a plateau $\Delta \theta = 2/N$~\cite{PezzePRL2009}. 
The Heisenberg limit $\Delta \theta = 1/N$ is achieved at $V_0 t = \pi/2$ and odd values of $N$.
For even values of $N$ the Heisenberg limit is reached at $V_0 t = \pi/2$ upon optimization of the rotation direction. 
It should be noted that Eq.~(\ref{FQ_OAT_1}) is not optimized over the rotation angle of the phase transformation. 
In this case, we have
\be \label{xiR_OAT_1}
\xiR^2 = \frac{N (\Delta \hat{J}_x)^2}{\mean{\hat{J}_z}^2} = (\cos V_0 t)^{-2(N-1)},
\ee
which is always larger than one, signaling the absence of spin squeezing orthogonal to the $y$ axis.
When optimizing over the rotation angle, 
{\it i.e.} replacing $\hat{U}_{1}$ with $\hat{U}_{1}^{\rm opt} = e^{-i \delta \hat{J}_n} e^{-i V_0 t \hat{J}_x^2}$ (and analogous for $\hat{U}_2^{\rm opt}$), 
where $n$ is the optimal rotation direction in the $x$-$y$ plane and $\delta$ the rotation angle, 
we obtain a larger QFI: 
\be \label{FQ_OAT_2}
F_Q\big[\ket{\psi_1}, \hat{J}_n \big] =  \max\big(F_Q^{(x,y)}, F_Q^{(z)}\big),
\ee
where
\be
F_Q^{(x,y)} = N + \frac{N(N-1)}{4} (A + \sqrt{A^2 + B^2}),
\ee
\be
F_Q^{(z)} = N^2 C - \frac{N(N-1)}{2}A,
\ee
$A = 1 - \cos^{N-2} 2 V_0 t$, $B = 4 \sin t \cos^{N-2}V_0 t$ and $C=1-\cos^{2(N-1)}V_0 t$.
The optimized spin squeezing is
\be \label{xiR_OAT_2}
\xiR^2 =  \frac{N (\Delta \hat{J}_\perp)^2}{\mean{\hat{J}_z}^2} = \frac{4+ (N-1) (A - \sqrt{A^2 + B^2})}{4 \cos^{2N - 2} V_0 t},
\ee
where $\perp$ is a direction perpendicular to $n$ and $z$.
The state is spin-squeezed for times $V_0 t \lesssim 1/\sqrt{N}$, while ENGSs are created for $V_0 t \gtrsim 1/\sqrt{N}$.
Dashed lines in Fig.~\ref{fig:dynamics}(a) [Fig.~\ref{fig:dynamics}(b)] show the sensitivity of the OAT model with $N=100$ particles 
obtained from the non-optimized [optimized] Loschmidt echo dynamics.

\begin{figure}[t!]
\includegraphics[width=0.48\textwidth]{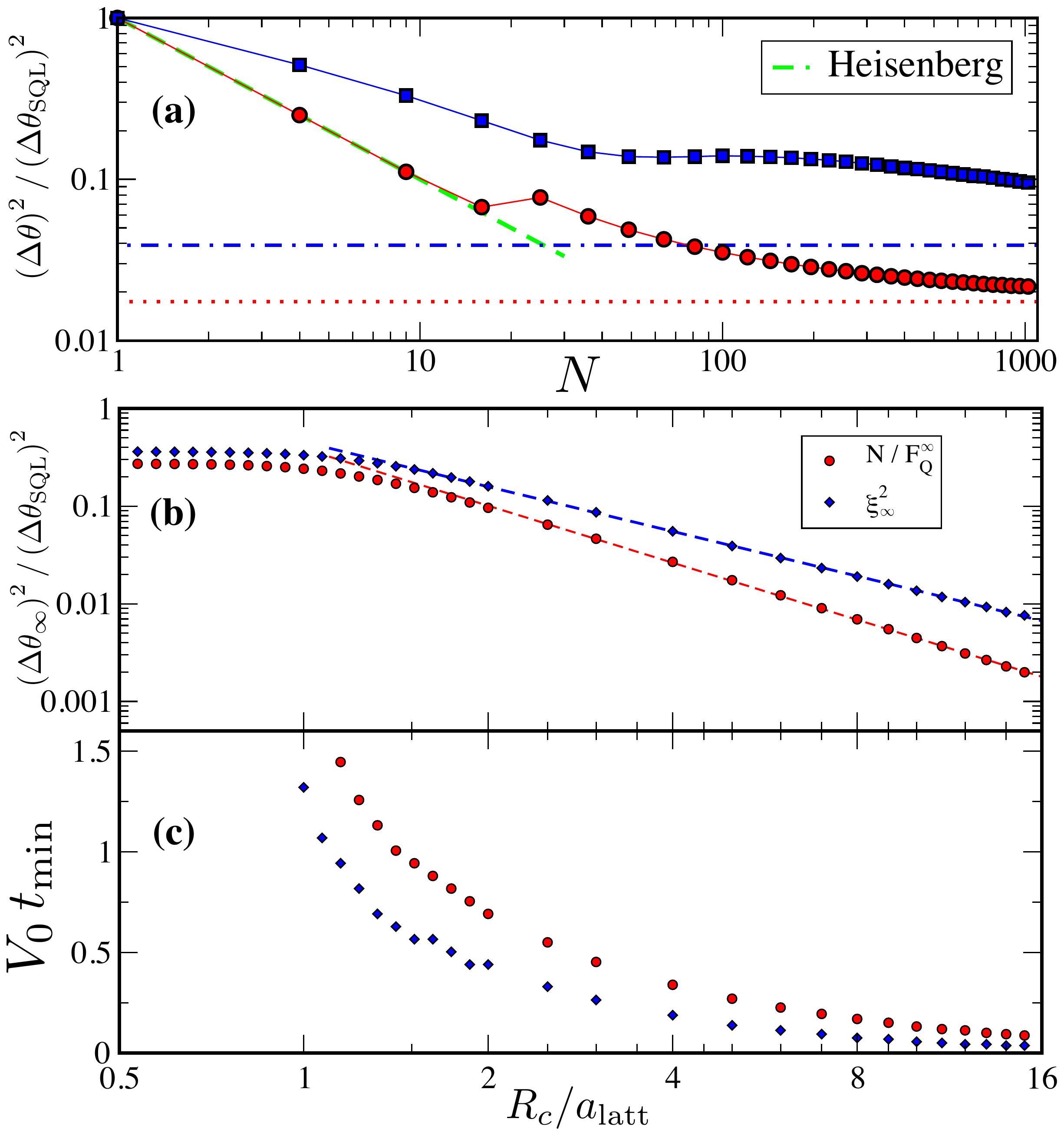} \\
\caption{(color online) 
(a) Comparison of phase sensitivities obtained via the inverse QFI (red dots) and spin squeezing (blue dots)
as a function of the number of particles in a 2D square optical lattice with $R_c = 5\, \text{a}_{\text{latt}}$.
Here the sensitivity is optimized with respect of evolution time and rotation direction. 
Solid lines are a guide for the eye.  
The dashed green line is the Heisenberg limit $\left(\Delta \theta\right)/\left(\Delta \theta_{\rm SQL}\right)^2=1/N$ achieved when 
the soft-core radius is larger than the maximum inter-particle distance.
Red dashed-dotted (blue dotted) is the limiting value of the inverse QFI $F_Q^\infty$ (spin squeezing $\xi_\infty^2$) for an infinite system.
(b) Minimum value of the inverse QFI and squeezing for an infinite two-dimensional system with varying soft-core radius. 
For the power law scaling we find: $ \xiR^2 \propto (Rc/a)^{-1.52}$ and $N/F_Q \propto (Rc/a)^{-1.94}$. 
(c) Time $V_0\, t_\text{min}$ where the minimum of the squeezing and the Quantum Fisher is obtained as a function of the soft-core radius.
}
\label{fig:number_p}
\end{figure}

Figure~\ref{fig:number_p}(a) shows the sensitivity (maximized over evolution time) achieved as a function of the number of particles in a 
two-dimensional setup
for $R_c=5\times \text{a}_{\text{latt}}$.
For small systems the minima of the inverse QFI scale as $N^{-2}$, reaching the Heisenberg limit as discussed above. 
For larger systems both the inverse QFI and spin squeezing display a shoulder due to finite size effects and then decrease.
In the thermodynamic limit the sensitivity is expected to scale linearly with the number of particles as faraway atoms uncorrelate
due to the finite range of the interaction potential.
In Fig.~\ref{fig:number_p}(b) we study more in detail such scaling for infinite two-dimensional systems as a function 
of the blockade radius. We observe a scaling of the sensitivity as a power law of the blockade radius, 
which then defines a characteristic entangling distance of close-by atoms. 
In Fig.~\ref{fig:number_p}(c) we display the optimal times at which the minima are obtained showing that they both diminish with
increasing blockade radius.
In all cases, ENGSs outperform the sensitivity achievable with spin-squeezed states even for a relatively small blockade radius. 
Notably this entanglement is fully exploited by the Loschmidt echo protocol.

%%%%%%%%%%%%%%%%%%%%%%%%%%%%%%%%%%%%%%%%%%%%%%%%%%%%%%%%
% figure 3
%%%%%%%%%%%%%%%%%%%%%%%%%%%%%%%%%%%%%%%%%%%%%%%%%%%%%%%%
%%%%%%%%%%%%%%%%%%%%%%%%%%%%%%%%%%%%%%%%%%%%%%%%%%%%%%%%%
%%%%%%%%%%%%%%%%%%%%%%%%%%%%%%%%%%%%%%%%%%%%%%%%%%%%%%%%%
%%%%%%%%%%%%%%%%%%%%%%%%%%%%%%%%%%%%%%%%%%%%%%%%%%%%%%%%%

%%%%%%%%%%%%%%%%%%%%%%%%%%%%%%%%%%%%%%%%%%%%%%%%%%%%%%%%
% figure 2
%%%%%%%%%%%%%%%%%%%%%%%%%%%%%%%%%%%%%%%%%%%%%%%%%%%%%%%%
\begin{figure}[t!]
%\begin{center}
\centering
\includegraphics[width=0.49\textwidth]{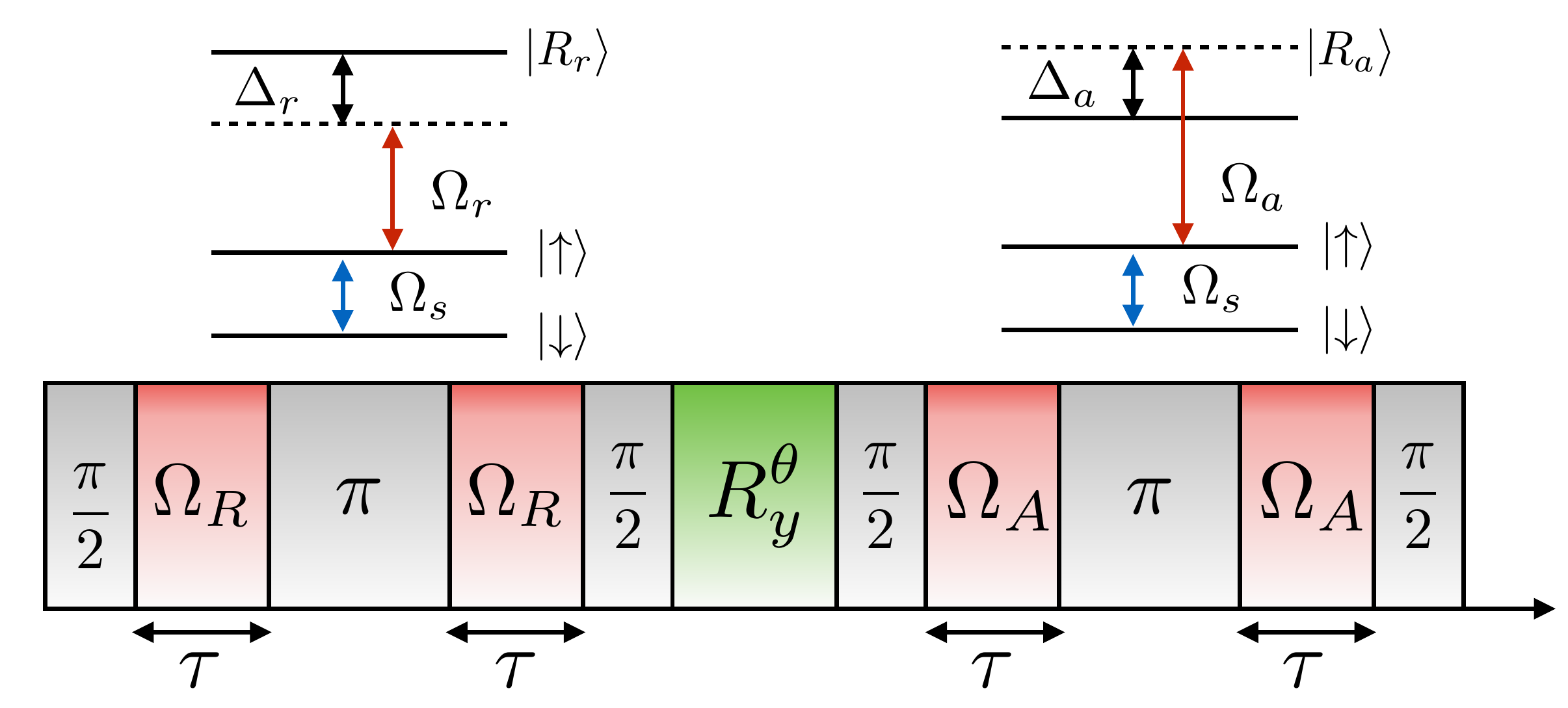}
\caption{(color online) 
Implementation of Loschmidt echo with Rydberg dressed atoms in lattices.
Upper panel: Level scheme with two lower energy levels that form an effective spin-$1/2$ system. Spin up is coupled
to a highly excited Rydberg state which displays repulsive (attractive) interactions in the first (second) part of the protocol.
Lower panel: Loschmidt echo protocol is implemented by two consecutive spin-echoes with a global spin rotation in the middle
by an angle $\theta$. Each spin-echo guarantees that inhomogeneous laser detunings decouple from the collective spin dynamics.
}
\label{fig:rydberg}
%\end{center}
\end{figure}
%%%%%%%%%%%%%%%%%%%%%%%%%%%%%%%%%%%%%%%%%%%%%%%%%%%%%%%%%
%%%%%%%%%%%%%%%%%%%%%%%%%%%%%%%%%%%%%%%%%%%%%%%%%%%%%%%%%
%%%%%%%%%%%%%%%%%%%%%%%%%%%%%%%%%%%%%%%%%%%%%%%%%%%%%%%%%

We further study here the possible experimental  implementation of the Loschmidt echo with Rydberg atoms in a lattice. 
One of the main motivation is that the interaction between Rydberg atoms trapped in a lattice may be a crucial strategy 
to create entangled useful states to increase the sensitivity of lattice clocks~\cite{LudlowRMP2015}. 
A key issue that we address here is how to invert, in practice, the sign of the interaction strength in the Hamiltonian~(\ref{LRI}) in order 
to close the echo protocol. 
The internal level structure of each atoms is represented in Fig.~\ref{fig:rydberg}.
The lower levels are two hyperfine states of an alkali atom or the two clock states of 
an alkaline earth atom. These two levels form an effective qubit. Operations are performed
by a laser field characterized by the Rabi frequency $\Omega_s$. 
The upper qubit state is then weakly admixed to a Rydberg state via a far-off resonant laser field with
coupling $\Omega_r$ ($\Omega_a$) and detuning $\Delta_r$ ($\Delta_a$) for repulsive (attractive) interactions. 
Notably, interaction between Rydberg atoms can be switched on and off almost instantaneously.

For the realization of the Loschmidt echo protocol we choose a repulsive (attractive) Rydberg level
in the first (second) part of the dynamics. The far off-resonant excitation 
allows to adiabatically eliminate the Rydberg state~\cite{Henkel2010, macri_2014},
at the cost of modifying the usual van der Walls interactions into effective soft-core inter-particle potentials  
$V_{ij}^{(a,r)}=\tilde C_6^{(a,r)}/((R_c^{(a,r)})^6+r_{ij}^6)$,
with $\tilde C_6^{(a,r)}=\frac{\Omega^4_{a,r}}{8\Delta^3_{a,r}} C_6^{(a,r)}$ the rescaled Van der Walls interaction coefficient 
for the attractive and repulsive Rydberg levels and $R_c^{(a,r)}=(C_6^{(a,r)}/2\Delta_{a,r})^{1/6}$ the soft-core radius.
Since kinetic terms (here we assume a deep optical lattice) and interactions among atoms in the ground state are negligible 
and the two Rydberg states are only weakly mixed to the lower qubit states 
the system can be regarded as an effective spin-$1/2$ system described by standard
Pauli operators:
$\hat \sigma_z^{(i)}=\left|g_i\right>\left<g_i\right|-\left|e_i\right>\left<e_i\right|$,
$\hat \sigma_x^{(i)}=\left|e_i\right>\left<g_i\right|+\left|g_i\right>\left<e_i\right|$, and
$\hat \sigma_y^{(i)}=i\, \left(\left|e_i\right>\left<g_i\right|-\left|g_i\right>\left<e_i\right|\right)$ 
and Hamiltonian:
\begin{equation} \label{ryd_ham}
H = \frac{\hbar\Omega_s}{2} \sum_{i=1}^N \hat \sigma_x^{(i)} +
\sum_{i<j}^N \frac{V_{ij}^{(a,r)}}{4}\hat \sigma_z^{(i)} \hat \sigma_z^{(j)}+
\sum_{i=1}^N \Delta_i \hat \sigma_z^{(i)},
\end{equation}
The implementation of the scheme can be done within the current experimental 
capabilities either with rubidium atoms or with alkaline-earth atoms like strontium or 
ytterbium atoms excited to Rydberg states using a sequence of two spin echoes which remove the effect of
inhomogeneous detuning as shown in Fig.~\ref{fig:rydberg}.
We limit here the discussion to a specific implementation with Rb atoms.
The qubit states can be chosen as the two hyperfine $\left|F=1\right>$ and $\left|F=2\right>$ states. 
Rydberg state $65\, D_{3/2}$ displays attractive interactions; 
following Ref.~\cite{Maucher2011} for an effective two-photon 
Rabi frequency $\Omega_a/2\pi = 0.5$~MHz via the $5P_{1/2}$ state and laser detuning 
$\Delta_a/2\pi=32$~MHz one gets an effective dressing potential with $R_c=5.9$~$\mu$m.
The strength of the potential is $W_a/\hbar = \frac{\Omega_a^4}{8\Delta_a^3} = 2\pi\cdot 0.238$~Hz and the
Rydberg state lifetime $\tau_{65_D} \simeq 280$~$\mu$s  \cite{Beterov2009} gets enhanced up to few seconds 
$\tilde \tau_{65_D}=\tau_{65_D}/\left(\Omega_a/2\Delta_a\right)^2 \gtrsim 4.5$~s.
For the second Rydberg state we choose the $80S_{1/2}$ which displays repulsive interactions. 
The requirement on this second state is to satisfy 
the condition $H_\text{r}=-H_\text{a}$ which implies that the soft-core radii in the two echo sequences need to be equal.
With $\Delta_r/2\pi = 50$~MHz and 
$\Omega_r/2\pi=0.75$~MHz we get an interaction strength $W_r/\hbar=2\pi \cdot 0.317$~Hz
combined with an even longer lifetime $\tau_{80_S} \sim 620$~$\mu$s, which gives $\tilde \tau_{80_S} \gtrsim 11$~s.
The soft-core interaction potential is certainly weak compared to the original Van der Walls interaction, however
the duration of the protocol does not exceed one tenth of the coherent lifetimes expected from these
implementation.
%A possible limitation of the present scheme is the decay of the intermediate state that can
%reduce the effective coherent dynamics to few hundreds of milliseconds making the scheme 
%feasible only for very large blockade radii $R_c\gtrsim 15\,a_\text{latt}$ and system sizes.
In a recent experiment the Munich group \cite{Zeiher2016} explored single photon excitation to anisotropic Rydberg 
P-states and demonstrated the feasibility of Rydberg dressing in optical lattices.
Coupling to states with P-symmetry ensures much higher interaction strengths ($\sim 1$~kHz)
and therefore a much faster implementation of the protocol.
Similarly single photon excitation from one of the clock states of alkaline-earth atoms may be a feasible alternative \cite{GilPRL2014}.

For a discussion of the most relevant detection errors we refer to the Munich setup. 
The reconstruction of the Loschmidt-echo
probability distribution $P_0(\theta)$ Eq.~(\ref{Eq.P0}) can be done by single (either spin-up or spin-down) 
or full spin resolution measurements (both spin-up and spin-down) 
as recently realized in \cite{Fukuhara2015} with in-situ Stern-Gerlach imaging. 
In both cases
it may happen that a tiny fraction of atoms 
$\epsilon$ ($\sim 1\%$) is lost during the measurement process. A lower bound
for the QFI due to this effect gives 
$F_Q^{\text{noise}}[\ket{\psi_1}, \hat{J}_n]\geq (1-\epsilon)F_Q[\ket{\psi_1}, \hat{J}_n]$ 
which shows that finite detection does not degrade the QFI significantly. Particle number fluctuations in the initial configuration
can be detrimental, especially when single-spin resolution is performed. 
Contrarily, when full spin resolution is employed,
sensitivity is not affected importantly. As an example, we computed that for an OAT system with 
initial configuration following a gaussian distribution with $N=100 \pm 7$, QFI is reduced by $2\%$. 
%This shows that postselection on the configurations with fixed particle number is not generically needed.

In conclusion, we have presented a versatile Loschmidt echo protocol for the creation and detection of ENGSs. 
It can be implemented in a variety of platforms, from ions to Rydberg atoms, from BECs to polar molecules. For the evolution of
a pure state the protocol saturates the QFI. % for generic interactions with arbitrary range and strength. 
In view of possible applications to lattice clocks, we have focused here on the implementation with Rydberg dressed atoms.
By choosing suitable Rydberg levels, 
it is possible to tune the interaction from attractive to repulsive while preserving the shape of the potential,  
and thus realize the Loschmidt echo protocol.
%and characterized the QFI and spin squeezing analytically via a mapping
%of the Hamiltonian to a spin-$1/2$ system. 
Even for system size larger than the typical interaction radius, the nonlinear dynamics generates %Gaussian spin-squeezed states 
ENGSs that are more useful (for metrological purposes) than the spin-squeezed states generated on relatively short time scales.
%while for longer times more useful   are generated. 
%We also showed that the results are solid with respect to uncertainties on the preparation of the 
%initial state and the exact inversion of the dynamics.
%Nevertheless, the Loschmidt echo techniques is quite general and can be exploited in different 
%state-creation schemes, including the more familiar one-axis twisting.
Applications of this protocols are within current experimental reach and they
may reveal important for the implementation of next-generation quantum technological devices.

{\it Acknowledgments}.
We thank C. Gross, D. Linnemann, T. Pohl, P. Schau\ss, A.M. Rey and J. Bollinger for useful discussions and insights.
T.M. acknowledges CNPq for support through the fellowship Bolsa de produtividade em 
Pesquisa n. 311079/2015-6.

\newpage

%\onecolumngrid

%\newpage

\end{document}